\documentclass[twocolumn]{aastex631}
\usepackage{amsmath}

\newcommand{\be}{\begin{equation}}
\newcommand{\ee}{\end{equation}}
\shorttitle{The Hubble Tension Revisited}
\shortauthors{M\"ortsell et al.}

\begin{document}

\title{The Hubble Tension Revisited:\\
Additional Local Distance Ladder Uncertainties}

\correspondingauthor{Edvard M\"ortsell}
\email{edvard@fysik.su.se}

\author{Edvard M\"ortsell}
\affiliation{Oskar Klein Centre, Department of Physics, Stockholm University\\Albanova University Center\\ 106 91 Stockholm, Sweden}

\author{Ariel Goobar}
\affiliation{Oskar Klein Centre, Department of Physics, Stockholm University\\Albanova University Center\\ 106 91 Stockholm, Sweden}

\author{Joel Johansson}
\affiliation{Oskar Klein Centre, Department of Physics, Stockholm University\\Albanova University Center\\ 106 91 Stockholm, Sweden}

\author{Suhail Dhawan}
\affiliation{Institute of Astronomy\\
University of Cambridge
Madingley Road \\ Cambridge CB3 0HA\\
United Kingdom}



\begin{abstract}
In a recent paper, we investigated possible systematic uncertainties related to the Cepheid color-luminosity calibration method and their influence on the tension between the Hubble constant as inferred from distances to Type Ia supernovae and the cosmic microwave background as measured with the Planck satellite. Here, we study the impact of other sources of uncertainty in the supernova distance ladder, including Cepheid temperature and metallicity variations, supernova magnitudes and GAIA parallax distances. Using Cepheid data in 19 
Type Ia supernova host galaxies from \citet{Riess:2016jrr},
anchor data from \citet{Riess:2016jrr, Riess_2019, Riess_2021} and a set of re-calibrated Milky Way Cepheid distances, we obtain $H_0 = 71.9\pm 2.2$ km/s/Mpc, $2.0\,\sigma$ from the Planck value. Excluding Cepheids with estimated color excesses $\hat{E}({\rm V} - {\rm I})=0.15$ mag to mitigate the impact of the Cepheid color-luminosity calibration, the inferred Hubble constant is $H_0 = 68.1\pm 2.6$ km/s/Mpc, removing the tension with the Planck value. 
\end{abstract}



\section{Introduction}
Cepheid stars are crucial in building up the distance ladder to Type Ia supernovae (SNIa) when estimating the Hubble constant $H_0$. To be used as standard candles, Cepheids need to be calibrated with respect to fact that
\begin{itemize}
    \item long period Cepheids are brighter,
    \item red Cepheids are dimmer,
    \item and Cepheids in high metallicity environments may be brighter. 
\end{itemize}
The Cepheid color-luminosity (C-L) correlation may be understood as a combination of extrinsic dust extinction and intrinsic temperature variations \citep{Pejcha_2012}. Given the difficulty in separating these effects, Cepheids are usually calibrated using a phenomenological approach where a parameter, $R$, corrects for both dust and intrinsic variations \citep{1982ApJ...253..575M}. The correction can be applied to observed colors, as by the SH0ES team in e.g. \citet{Riess:2016jrr}, or estimated color excesses, as in \citet{Follin_2018}, derived by subtracting a model for the mean intrinsic Cepheid color. Regardless of which method is employed, it will by necessity correct for both extrinsic and intrinsic color variations since the observed color and estimated color excess only differ by a linear term depending on the Cepheid period. When assuming a global value for $R$, the choice of calibration method is of minor importance since only the difference between Cepheid magnitudes in anchor and SNIa host galaxies matters for the $H_0$ inference. When allowing for the calibration parameter to vary between galaxies, motivated by the observed variation in dust properties \citep{mortsell2021hubble}, the choice of calibration can potentially make  a difference. As demonstrated in \citet{mortsell2021hubble}, using Wesenheit H-band magnitudes calibrated with respect to the observed color $R_{\rm W}({\rm V} - {\rm I})$ where a global $R_{\rm W}$ is allowed to vary gives 
$H_0=73.2\pm 1.3$ (in units of km/s/Mpc implied from now on), whereas calibrating using estimated color excesses with individual galactic $R_{\rm E}\,\hat{E}\,({\rm V} - {\rm I})$ yields $H_0=73.9\pm 1.8$, a $3.4\,\sigma$ tension with the value inferred from the cosmic microwave background (CMB) observations with the Planck satellite \citep{Planck2020}. 
Arguably being closest to a physical Cepheid model as presented in \cite{Pejcha_2012}, throughout the paper, we will use the color excess calibration as our default method for investigation\footnote{The SH0ES team have presented an updated Hubble constant value of $H_0=73.04\pm 1.04$ ($4.9\,\sigma$ Planck tension) using an, yet not publicly available, expanded SNIa and Cepheid data set \citep{Riess:2021jrx}. As in \citet{mortsell2021hubble}, we here make use of the data from \citet{Riess:2016jrr, Riess_2019, Riess_2021} and \citet{Breuval_2020}.}. 

We first point out that the Cepheid color excess distributions have substantial variations between galaxies, and that the inferred $H_0$ is sensitive to the introduction of color cuts in the data. Next, we investigate the impact of Cepheid temperature variations on the inferred $H_0$ uncertainty and discuss how systematic effects connected to the metallicity calibration may shift the inferred $H_0$ and increase the error budget. We express concerns about the reliability of the Milky Way (MW) Cepheid parallax distance calibration and show that the local Cepheid calibrated SNIa magnitude uncertainties may be underestimated. Finally, we combine the different systematic effects using Monte Carlo simulations and derive constraints on the Hubble constant. 

\section{Method and Data}

\subsection{Cepheid Calibration}
We use the Hubble Space Telescope (HST) near-infrared flux (${\rm H}={\rm F160W}$), color calibrated using optical (${\rm V}={\rm F555W}$ and ${\rm I}={\rm F814W}$) data, to derive
\be\label{eq:whevi} 
m_{\rm H}^{\rm W} \equiv m_{\rm H} - R_{\rm E}\,\hat{E}\,({\rm V} - {\rm I}).
\ee
Here, $\hat{E}\,({\rm V} - {\rm I})$ represents a proxy for the color excess  $E\,({\rm V} - {\rm I})\equiv A_{\rm V}-A_{\rm I}=({\rm V} - {\rm I})-({\rm V} - {\rm I})_0$, with $({\rm V} - {\rm I})_0$ the intrinsic Cepheid color.
It is obtained by subtracting an estimate of the mean intrinsic colors, $\langle{\rm V} - {\rm I}\rangle_0$ from the observed colors. We use mean intrinsic colors from \citet{Tammann_2011}, including the quoted uncertainties and a $0.075$ dispersion in the mean intrinsic Cepheid color between galaxies, from the difference between Large Magellanic Cloud (LMC) and MW Cepheids, see Table~\ref{tab:Tamman}. 
\begin{table}[!ht]
  \centering
  \caption{Estimated mean intrinsic colors, $\langle{\rm V} - {\rm I}\rangle_0$ from \citet{Tammann_2011}. Here, $[{\rm P}]\equiv \log P -1$ where $P$ is the Cepheid period measured in days.} 
  \label{tab:Tamman}
  \begin{tabular}{|c|c|}
    \hline
    $\langle{\rm V} - {\rm I}\rangle_0$ & Galaxy\\
    \hline
    $(0.753\pm 0.023) + (0.256\pm 0.017)[{\rm P}]$ & MW\\
    $(0.675\pm 0.02) + (0.201\pm 0.017)[{\rm P}]$ & LMC, $[{\rm P}]<-0.1$\\
    $(0.676\pm 0.038) + (0.345\pm 0.024)[{\rm P}]$ & LMC, $[{\rm P}]>-0.1$\\
    \hline
  \end{tabular}
\end{table}
These colors are in good agreement with results in \citet{Pejcha_2012}. 
We fit for the values of $R_{\rm E}$ that minimize the scatter in $m_{\rm H}^{\rm W}$.

The Wesenheit magnitude of the $j$th Cepheid in the $i$th galaxy, including the anchor galaxies MW, NGC 4258 and the Large Magellanic Cloud (LMC), is modeled as
\be\label{eq:mHW}
m_{{\rm H},i,j}^{\rm W }=\mu_i+M_{\rm H}^{\rm W}+b_{\rm W}[{\rm P}]_{i,j}+Z_{\rm W} [{\rm M}/{\rm H}]_{i,j},
\ee
with $[{\rm M}/{\rm H}]_{i,j}$ the Cepheid metallicity, $M_{\rm H}^{\rm W}$ the absolute Cepheid magnitude normalized to a period of $P=10$ days and Solar metallicity and $\mu_i$ the distance modulus to the $i$th galaxy. We use separate P-L relations for short and long period Cepheids 
\be
b_{\rm W}[{\rm P}]_{i,j}\rightarrow b^{\rm s}_{\rm W}[{\rm P}]^{\rm s}_{i,j}+ b^{\rm l}_{\rm W}[{\rm P}]^{\rm l}_{i,j},
\ee
where $[{\rm P}]^{\rm s}_{i,j}=0$ for Cepheids with periods $>10$ days and  $[{\rm P}]^{\rm l}_{i,j}=0$ for Cepheids with periods $<10$ days.

\subsection{Type Ia Supernovae}
The apparent SNIa B-band peak magnitude in the $i$th host, corrected for the width-luminosity and C-L relations using the SALT2 model \citep{2007A&A...466...11G}, is modeled by
\be 
m_{{\rm B},i}=\mu_i+M_{\rm B}.
\label{eq:snmag} 
\ee
The errors on $m_{{\rm B},i}$ include the fitting uncertainty and a $0.1$ magnitude contribution to take into account the intrinsic SNIa dispersion added in quadrature \citep{Riess:2016jrr,mortsell2021hubble}.

\subsection{Data and Parameter Fitting}
For Cepheids in M31 and beyond, we use data from Table~4 in \citet{Riess:2016jrr}. For Cepheids in the LMC, we use data in Table~2 in \citet{Riess_2019}. Data for MW Cepheids, including GAIA parallax measurements are from Table~1 in \citet{Riess_2021}.

20 double eclipsing binaries (DEBs) observed using long-baseline near-infrared interferometry give a distance to the LMC of $\mu_{\rm LMC}=18.477\pm 0.0263$ mag \citep{Paczynski:1996dj,Pietrzy_ski_2019,Riess_2019}.
Observations of mega-masers in Keplerian motion around its central super massive black hole give a distance to NGC 4258 of $\mu_{\rm N4258}=29.397\pm 0.032$ mag \citep{Reid_2019}.

Type Ia SN B-band magnitudes are from Table~5 in \citet{Riess:2016jrr}, derived using version 2.4 of SALT II \citep{Betoule_2014}.

Given the observed Cepheid magnitudes $m_{\rm H}$, color excesses $\hat{E}\,({\rm V} - {\rm I})$, periods $[{\rm P}]$, metallicities $[{\rm M}/{\rm H}]$, together with the SNIa magnitudes $m_{\rm B}$, the anchor distances $\mu_k$ and the MW Cepheid parallaxes $\pi$, we fit simultaneously for $R_{\rm E}$, $b_{\rm W}$, $Z_{\rm W}$, the host galaxy distances $\mu_i$, the anchor distances $\mu_k$, the GAIA parallax offset $zp$, the Cepheid absolute magnitude $M_{\rm H}^{\rm W}$ and the SNIa absolute magnitude $M_{\rm B}$. For linear parameters, the fit can be made analytically. The exception is the $R_{\rm E}$; since the uncertainties in the observed Cepheid colors are non-negligible, the uncertainties in the derived Wesenheit magnitudes $m_{\rm H}^{\rm W}$ will depend on the values of $R_{\rm E}$, requiring a non-linear treatment of these parameters using  Markov chain Monte Carlo (MCMC) techniques, see \citet{mortsell2021hubble}.

\subsection{Default Results for Color excess Calibration 
\texorpdfstring{$R_{\rm E}\,\hat{E}({\rm V} - {\rm I})$}{}}
Using similar assumption as in \citet{mortsell2021hubble}, allowing for $R_{\rm E}$ to vary between galaxies, with weak flat prior constraints on their values $R_{\rm E}=[0,1]$, we obtain $H_0=73.9\pm 1.5$, in $4.0\,\sigma$ tension with the Planck value. For a fixed global value of $R_{\rm E}=0.386$, following \citet{Riess:2016jrr}, we obtain $H_0=73.0\pm 1.3$ ($4.0\,\sigma$).
These are the default results to which we will relate the impact of other systematic effects, the latter having the advantage of only including linearly fitted parameters, substantially reducing the computational cost when comparing different scenarios. Note that the error bars do not yet include the intrinsic Cepheid color scatter, see Section~\ref{sec:results}.

\section{Color excess distributions and cuts}
\begin{figure}[!ht]
    \centering
	\includegraphics[width=1\linewidth]{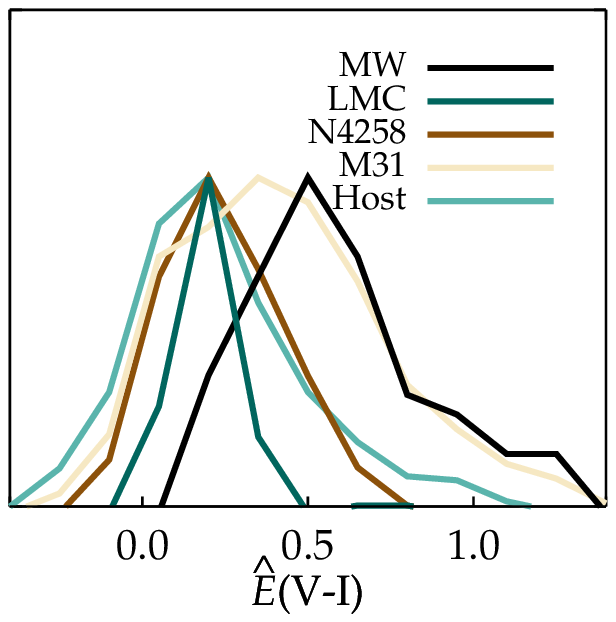}
	\caption{Normalized distributions of estimated Cepheid color excesses $\hat{E}({\rm V} - {\rm I})=({\rm V} - {\rm I})-\langle{\rm V} - {\rm I}\rangle_0$, using HST optical $(V-I)$ data, and mean intrinsic colors $\langle{\rm V} - {\rm I}\rangle_0$ from \citet{Tammann_2011}. The mean color excesses $\langle\hat{E}({\rm V} - {\rm I})\rangle$ are for the MW $0.69$, for the LMC $0.28$, for NGC 4258 $0.33$, for M31 $0.50$ and for the SNIa host galaxies $0.31$. 
	\label{fig:EVIhistnorm}}
\end{figure}

\begin{figure}[!ht]
    \centering
	\includegraphics[width=1\linewidth]{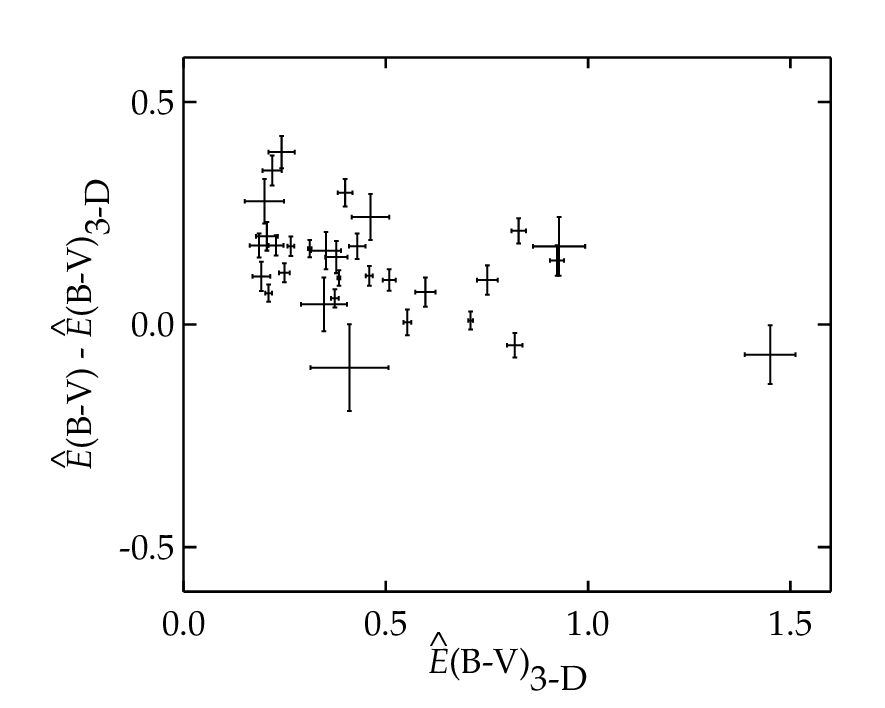}
	\caption{The ${\rm B} - {\rm V}$ estimated Cepheid color excess $\hat{E}({\rm B} - {\rm V})$ compared to the predicted dust extinction based on three dimensional (3-D) dust maps, $\hat{E}({\rm B} - {\rm V})_{\rm 3-D}$, estimated following \citet{2018JOSS....3..695G}.
	\label{fig:PanSvsCeph2}}
\end{figure}

As discussed in \citet{mortsell2021hubble}, the fact that Cepheid colors and periods are correlated \citep{Tammann_2011} may cause color selection effects related to the fact that longer period Cepheids are brighter. 
From Figure~\ref{fig:EVIhistnorm}, showing the different distributions of the color excess $\hat{E}({\rm V} - {\rm I})$, large variations between the different galaxies employed in the Cepheid distance calibration are evident. 
This fact, possibly attributed to selection effects, may introduce systematic errors when comparing the derived Cepheid distances.
MW are systematically redder compared to other anchors and SNIa host galaxies, and also have a strong correlation of long period Cepheids having larger color excesses. For each individual MW Cepheid, we compare the estimated color excess with three dimensional dust maps based on GAIA, Pan-STARRS 1 and 2MASS data where applicable \citep{Green_2019,2018JOSS....3..695G}. The generally good agreement as shown in Figure~\ref{fig:PanSvsCeph2} confirms that the majority of the color excess is due to interstellar dust extinction, but with room for contributions from circumstellar dust and intrinsic temperature variations.

Given that the statistical uncertainty contribution to $H_0$ is dominated by SNIa magnitude errors, one can afford to make rather severe cuts in the Cepheid data in order to minimize the impact of systematic effects, including color cuts to try to homogenize the different color excess distributions between galaxies.

\begin{figure}[!ht]
    \centering
	\includegraphics[width=1\linewidth]{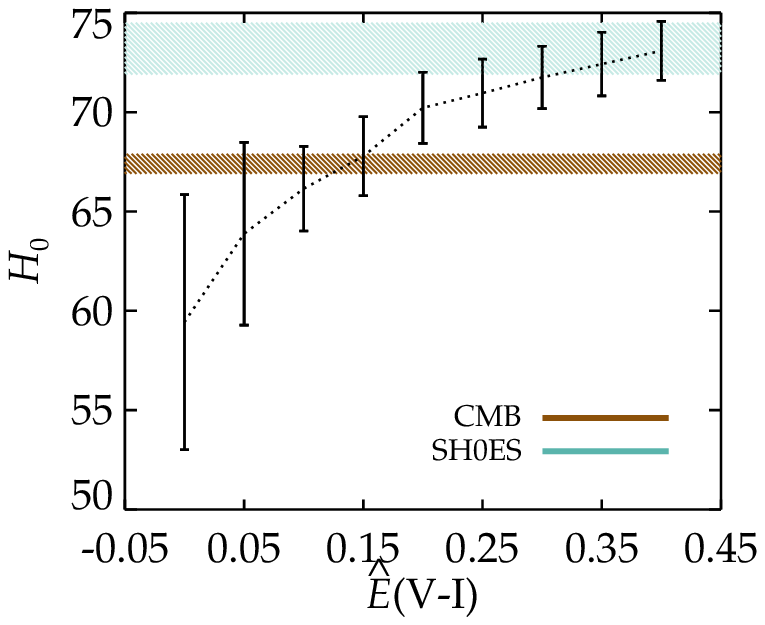}
	\caption{Fitted $H_0$ as a function of the cut in $\hat{E}({\rm V} - {\rm I})$ for individually fitted $R_{\rm E}$ with a weak prior. The error bars do not include intrinsic color and metallicity systematic errors.
	\label{fig:h0vsevicut_emcee}}
\end{figure}

Here, we investigate the impact on $H_0$ of mitigating the color calibration impact by removing the reddest Cepheids, most susceptible to dust extinction. 
In Figure~\ref{fig:h0vsevicut_emcee}, we show the fitted $H_0$ as a function of the cut in $\hat{E}({\rm V} - {\rm I})$ we apply, when calibrating using color excesses for individual $R_{\rm E}$ in the range $[0,1]$. We note that the size of the error bars do not significantly increase unless $\hat{E}({\rm V} - {\rm I})_{\rm max}\lesssim 0.1$ mag, and that the inferred $H_0$ drifts towards the Planck value when cutting out redder Cepheids. 

The reason for $H_0$ to decrease when removing red Cepheids is that even after applying the color calibration of equation~\ref{eq:whevi}, in individual galaxies there are still left-over correlations between the $m_{\rm H}^{\rm W}$ and $\hat{E}({\rm V} - {\rm I})$. This will be the case also when allowing for individual galactic $R_{\rm E}$ since the fitted value will not only depend on the slope of the magnitude-color diagram, but also on its normalization. After color calibration, $m_{\rm H}^{\rm W}$ and $\hat{E}({\rm V} - {\rm I})$ tend to be anti-correlated in SNIa host galaxies and correlated in anchor galaxies, see Table~\ref{tab:pearson}. Removing Cepheids with large $\hat{E}({\rm V} - {\rm I})$ therefore increases the mean $m_{\rm H}^{\rm W}$ in SNIa host galaxies and decreases the mean $m_{\rm H}^{\rm W}$ in anchor galaxies. As discussed in \citet{mortsell2021hubble}, both of these effects decrease the inferred $H_0$. The exception is MW Cepheids that have $m_{\rm H}^{\rm W}$ and $\hat{E}({\rm V} - {\rm I})$ 
with slightly stronger anti-correlation than SNIa hosts. However, the fact that the MW Cepheid color excess distribution is heavily biased toward high values means that when applying a cut in $\hat{E}({\rm V} - {\rm I})\lesssim 0.3$ mag, too few MW Cepheid will remain to contribute appreciably to the inferred $H_0$. 

\begin{table}[!ht]
  \centering
  \caption{The Pearson correlation coefficient, $\rho$ between $m_{\rm H}^{\rm W}$ and $\hat{E}({\rm V} - {\rm I})$ for $R_{\rm E}=0.386$.} 
  \label{tab:pearson}
  \begin{tabular}{|c|c|}
    \hline
    Galaxies & $\rho [m_{\rm H}^{\rm W},\hat{E}({\rm V} - {\rm I})]$\\
    \hline
    MW & $-0.27$\\
    LMC & $0.49$\\
    N4258 & $0.08$\\
    SNIa hosts& $-0.13$\\
    \hline
  \end{tabular}
\end{table}

Applying $\hat{E}({\rm V} - {\rm I})<0.15$ mag, motivated by this being the cut where statistical Cepheid uncertainties are still subdominant and we expect dust extinction not to dominate the Cepheid excess colors, we obtain $H_0 = 67.8\pm 2.0$ for individually fitted $R_{\rm E}=[0,1]$. Assuming a fixed global value of $R_{\rm E}=0.386$, closer in line with the analysis in \citet{Riess:2016jrr}, we obtain $H_0 = 68.0\pm 2.1$ for the same color cut. As evident from Figure~\ref{fig:EVIhistnorm}, this cut effectively removes all MW Cepheids. 

\section{Temperature variations}\label{sec:tempvar}
Given our inability to separate dust and Cepheid temperature color effects, we will here treat the temperature magnitude and color variations as an additional source of uncertainty, following the physical Cepheid model from \citet{Pejcha_2012}. Denoting the mean temperature deviation for a specific Cepheid by $\tau$, the absolute X-band magnitude will shift according to
\be
M_{\rm X}=\langle M_{\rm X}\rangle-2.5\beta_{\rm X}\tau,
\ee
where $\langle M_{\rm X}\rangle$ is the mean absolute magnitude at the specific period, color and metallicity. $\beta_{\rm X}$ for a large range of filters are observationally constrained in \citet{Pejcha_2012}, where also temperature variations are estimated to $\sigma_\tau=0.02$ given the width of the Cepheid instability strip. 
The intrinsic color (ic) excess will accordingly be given by 
\be
\hat{E}_{\rm ic}({\rm V} - {\rm I})=-2.5(\beta_{\rm V}-\beta_{\rm I})\tau.
\ee
For each Cepheid, we randomly assign a temperature from $\tau=0\pm \sigma_\tau$ and adjust the H-magnitude and color excess according to 
\begin{align}
  m_{\rm H}&\rightarrow m_{\rm H}+2.5\beta_{\rm H}\tau \nonumber \\
  \hat{E}({\rm V} - {\rm I})&\rightarrow \hat{E}({\rm V} - {\rm I})-\hat{E}_{\rm ic}({\rm V} - {\rm I}),
\end{align}
assuming $\beta_{\rm H}=1.72, \beta_{\rm I}=3.39$ and $\beta_{\rm V}=5.14$. The induced impact on the inferred $H_0$ is estimated generating random Monte Carlo samples, see Section~\ref{sec:results}. Note that it makes sense treating the temperature color excess as noise for the dust color excess since the former is subdominant, with $\hat{E}_{\rm ic}({\rm V} - {\rm I})\lesssim 0.1$ mag.

\section{Metallicities}\label{sec:metallicities}
We correct for possible metallicity effects on the luminosity using $\delta m_{\rm H}^{\rm W } = Z_{\rm W} [{\rm M}/{\rm H}]$, where the bracket is a shorthand notation for 
\begin{align}
[{\rm M}/{\rm H}]&\equiv \log\left[\frac{({\rm M}/{\rm H})}{({\rm M}/{\rm H})_\odot}\right]=\log ({\rm M}/{\rm H})-\log ({\rm M}/{\rm H})_\odot\nonumber \\
&=\Delta \log({\rm M}/{\rm H}).
\end{align}
For the default case with $R_{\rm E}=0.386$, we infer $Z_{\rm W}=-0.21\pm 0.05$.

In \cite{Riess:2016jrr}, individual Cepheid metallicities are estimated using $Z=12+\log ({\rm O}/{\rm H})$. For LMC Cepheids, a common value of $\Delta\log ({\rm O}/{\rm H})=-0.25$ was used. In \cite{Riess_2019}, this was updated to  $[{\rm Fe}/{\rm H}]=-0.3$, also used in \citet{mortsell2021hubble}, referring to \cite{Riess:2016jrr} having  $[{\rm Fe}/{\rm H}]=-0.25$ indicating the identification $[{\rm Fe}/{\rm H}]=\Delta\log ({\rm O}/{\rm H})$. 
However, since
\be
[{\rm O}/{\rm H}]=[{\rm O}/{\rm Fe}]+[{\rm Fe}/{\rm H}],
\ee
this is only true if $[{\rm O}/{\rm Fe}]=0$. As stated in \citet{Israelian_1998}, studies agree that [{\rm O}/{\rm Fe}] increase when [{\rm Fe}/{\rm H}] decrease from 0 to $-1$. Approximating this dependence to be linear, $[{\rm O}/{\rm Fe}]=k[{\rm Fe}/{\rm H}]$, we can write
\be
[{\rm O}/{\rm H}]=k[{\rm Fe}/{\rm H}]+[{\rm Fe}/{\rm H}]=[{\rm Fe}/{\rm H}](1+k).
\ee
Figure 3 in \cite{Amarsi_2015} and figure~20 in \cite{Luck_2018} suggest that $k\sim-0.5$, with an uncertainty of order $\sigma_k=0.25$. 

Subsequently, in \citet{Romaniello:2021vht} iron and oxygen abundances of of 89 Cepheids in the LMC yielded an updated metallicity of $[{\rm M}/{\rm H}]=-0.4079\pm 0.003$, which we will use as our default value for Cepheids in the LMC when deriving our final results in Section~\ref{sec:results}. 
When estimating the Cepheid metallicity from iron and oxygen abundances, state-of-art compilations differ by $\sim 0.15$ in the difference $Z_\odot-Z^{\rm Fe}_\odot$ \citep{Vagnozzi_2019}. In \cite{von_Steiger_2015}, a value of ${\rm H}/{\rm O}=1500\pm 300$ is given, corresponding to $Z_\odot=8.824\pm 0.087$ which we will use as our default value, with an $0.15$ dispersion to take into account the uncertainty between Cepheid metallicities estimated from oxygen and iron abundances.

\subsection{Impact on \texorpdfstring{$H_0$}{}}
The derived $H_0$ decreases slightly with decreasing the LMC metallicity. For a fixed value $R_{\rm E}=0.386$, shifting $\Delta\log ({\rm O}/{\rm H})=-0.3\rightarrow -0.4079$ decreases the inferred Hubble constant with $\delta H_0\sim 0.26$.

The inferred $H_0$ depends on $k$ as $\delta H_0/\delta k\sim -0.7$, and changing from $k=0$ to $k=-0.5$, we obtain $H_0=73.3\pm 1.3$ for $R_{\rm E}=0.386$.
A constant systematic shift in the Cepheid metallicities as inferred from iron and oxygen abundances can be parametrized by changing the assumed solar oxygen abundance. An increase $\delta Z_\odot$ will increase $H_0$ with $\delta H_0/\delta Z_\odot \sim 5$. Shifting  $Z_\odot=8.824\rightarrow 8.674$, for $R_{\rm E}=0.386$ we get $H_0=71.9\pm 1.4$ ($3.1\,\sigma$ tension).
The full impact on the inferred $H_0$ when allowing also for varying galactic $R_{\rm E}$ is estimated using Monte Carlo techniques in Section~\ref{sec:results}.

\section{Milky Way Parallax Uncertainties}\label{sec:mwparuncert}
Trigonometric parallaxes potentially provide the most direct calibration of the Cepheid absolute magnitude, $M_{\rm H}^{\rm W}$.
We use data from \citet{Riess_2021}, with 68 MW Cepheids having estimated GAIA parallaxes. As described in \citet{Lindegren_2021}, there are systematic errors in the published GAIA parallax values, with the parallax bias depending on, for example, the magnitude, color and angular position of the source. Although the GAIA team provide tentative expressions for the parallax correction, given that these corrections are uncertain for sources as bright as the MW Cepheids used in this study, here, as well as in \citet{Riess_2021}, we will allow for the possibility of residual parallax biases to correct for. Possible systematic GAIA Cepheid parallax uncertainties were investigated using globular clusters in \citet{2021MNRAS.505.5978V} and \citet{2021A&A...649A..13M}.
Based on MW Cepheid parallax uncertainties and metallicity effects, in \citet{2022ApJ...927....8O} and references therein, it was also concluded that the uncertainty in the derived Cepheid distance scale may be underestimated, suggesting a systematic error floor of $\sim 3\,\%$. 

For the $j$th Cepheid in the MW, 
\be 
m_{{\rm H},j}^{\rm W}=\mu_j+M_{\rm H}^{\rm W}+b_{\rm W}[{\rm P}]_{j}+Z_{\rm W} [{\rm M}/{\rm H}]_{j}.
\ee
where the distance modulii for each Cepheid is estimated using GAIA parallaxes, $\pi$, according to 
\be
\pi_j + zp=10^{-0.2(\mu_j-10)},
\ee
where $zp$ is a residual parallax calibration offset that we fit for together with $M_{\rm H}^{\rm W}, b_{\rm W}$ and $Z_{\rm W}$ by writing
\begin{align}
\mu_j&=10-\frac{5}{\ln 10}\left[\ln\pi+\ln\left(1+\frac{zp}{\pi}\right)\right]\nonumber \\& = 10-\frac{5}{\ln 10}\left[\ln\pi+\frac{zp}{\pi}+\mathcal O \left(\frac{zp}{\pi}\right)^2\right],
\end{align} 
effectively transforming $zp$ into a linear parameter, and
\begin{align}\label{eq:mpi}
m_{{\rm H},j}^{\rm W}-10+\frac{5}{\ln 10}\ln\pi& =M_{\rm H}^{\rm W}+b_{\rm W}[{\rm P}]_{j}\nonumber \\
&+Z_{\rm W} [{\rm M}/{\rm H}]_{j}-\frac{5}{\ln 10}\frac{zp}{\pi}.
\end{align} 
Higher order terms, $\mathcal O(zp/\pi)^2$, are small and corrected for in an iterative manner. So far, with the exception that we fit for $zp$ simultanously with all other parameters, this is similar to the approach in \citet{Riess_2021}. For the default case where we calibrate an individual extinction law for each galaxy in the interval $R_{\rm E}=[0,1]$, we obtain $zp=-18.9\pm 6.1\, \mu{\rm as}$. For a fixed global value of $R_{\rm E}=0.386$, $zp=-16.9\pm 5.1\, \mu{\rm as}$.

\begin{figure}[ht]
    \centering
	\includegraphics[width=1\linewidth]{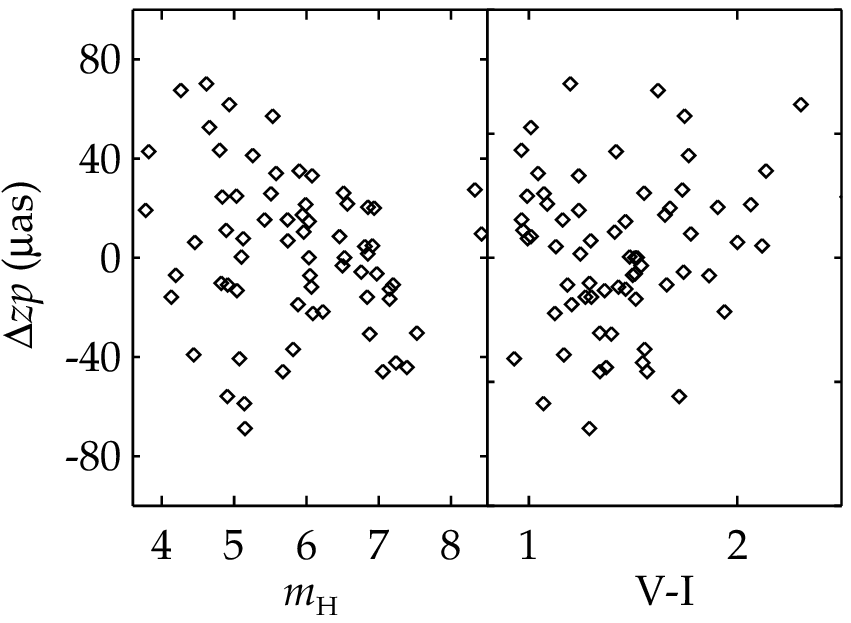}
	\caption{{\em Left panel:} Parallax offset residual vs $H$-band magnitude for MW Cepheids, with a negative Pearson correlation of $-0.22$. {\em Right panel:} Parallax offset residual vs ${\rm V}-{\rm I}$-color for MW Cepheids with a positive Pearson correlation $0.15$.
	\label{fig:MWres_full}}
\end{figure}

After correcting for a constant residual parallax offset in the GAIA data, we may have a residual that correlates with magnitude and color, see figure~\ref{fig:MWres_full}.
For the magnitude, we have a negative Pearson correlation of $-0.22$ and for the color a positive Pearson correlation $0.15$.
We therefore also add corrections of the form
\be
\pi\rightarrow \pi+zp_1+zp_2\cdot m_{\rm H} + zp_3\cdot ({\rm V} - {\rm I}).
\ee
Using the MW as the sole anchor galaxy with a fixed $R_{\rm E}=0.386$, adding $zp_2$ increases the Hubble constant from $H_0=73.8\pm 1.5$ to $H_0=77.0\pm 1.9$. Adding also $zp_3$ gives $H_0=72.5\pm 1.6$. 
The fact that a change in the parallax offset parameterization can induce a shift of $\delta H_0 =4.5$ introduces doubt about the uncertainty estimates. 

An interesting possibility for calibrating MW Cepheid distances circumventing problems connected to their bright nature and variability is to employ Cepheids for which the distance can be estimated from the parallax of spatially resolved
companions or their host open cluster. In \citet{Breuval_2020}, a sample of 36 MW Cepheids is constructed in this way, and assuming a fixed value for the GAIA calibration off-set, it was noted that the inferred $H_0$ is decreased comparing to the case of parallaxes measured from the Cepheids themselves. Here, we consider the impact of using the MW Cepheid sample from \citet{Breuval_2020} when performing the full simultaneous parameter distance ladder fit, including the GAIA parallax off-set and assuming $R_{\rm E}=0.386$. The inferred Hubble constant, $H_0=72.5\pm 1.4$ ($3.5\,\sigma$ Planck tension), is slightly decreased compared to using the Cepheids in \citet{Riess_2021}, with slightly larger uncertainties given the smaller MW Cepheid sample size. 
The value is very similar to the result obtained when skipping MW Cepheids altogether in which case $H_0 = 72.7 \pm 1.4$ ($3.5\,\sigma$), before taking other systematic uncertainties into account.

When allowing for individually fitted galactic $R_{\rm E}$, we obtain 
$H_0 = 71.3 \pm 1.6$ ($2.3\,\sigma$) using the \citet{Breuval_2020} MW Cepheid sample, again neglecting additional systematic effects. 

\section{Supernova Magnitude Uncertainties}\label{sec:snmag}
\begin{figure}[ht]
    \centering
	\includegraphics[width=\linewidth]{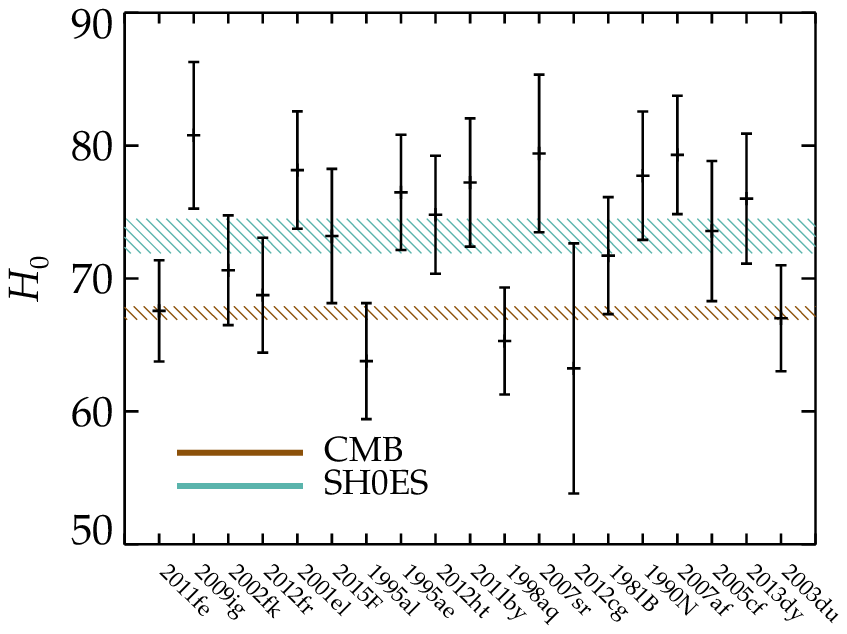}
	\caption{Fitting individual SNIa values of $H_0$ with a fixed global value of $R_{\rm E}=0.386$.
	\label{fig:indH0_intcol}}
\end{figure}

If we instead of fitting for a global absolute SNIa magnitude, fit for individual $M_{\rm B}$, we obtain one estimate of $H_0$ for each SNIa,
see figure~\ref{fig:indH0_intcol} for the case of a fixed global $R_{\rm E}=0.386$.
Apart from noting that only a subset of SNIae are in significant tension with the CMB result for $H_0$, we also note that the scatter is slightly larger than expected given the individual error bars, possibly indicating underestimated SNIa magnitude uncertainties. Taking the full correlation between the inferred $M_{\rm B}$ into account, the SNIa error bars should be increased by a factor of $1.22$ in order for the $\chi^2/{\rm dof}=1$ assuming the individual $M_{\rm B}$ have their origin in a common value. Increasing the $\sigma(M_{\rm B})$ by this factor, for the default case we obtain $H_0 = 73.0\pm 1.5$, decreasing the Planck tension slightly to $3.6\,\sigma$. Fitting galactic individual $R_{\rm E}=[0,1]$ increases the $H_0$ scatter and the SNIa errors bars need to be increased by by a factor of $1.37$. Fitting for a global value Hubble constant for default parameter values, gives $H_0 = 73.9\pm 1.8$ ($3.5\,\sigma$ tension). For a set of more restricted priors, with flat priors $R_{\rm E}=[0.15,0.8]$ and Gaussian priors $R_{\rm E}=0.48\pm 0.1$ \citep[see][]{mortsell2021hubble},
the corresponding SNIa error factors are $1.17$ and $1.23$ respectively.

\subsection{Impact of Type Ia Supernova Data Set}
\begin{figure}[ht]
    \centering
	\includegraphics[width=\linewidth]{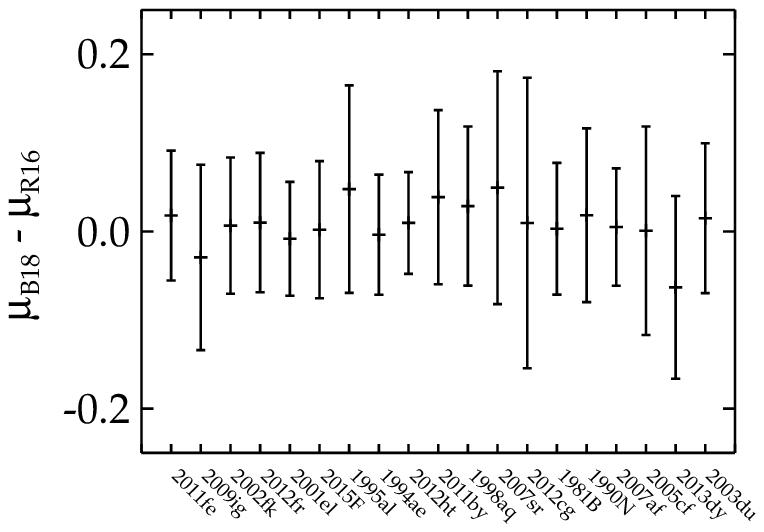}
	\caption{Comparing the distance modulii for SNIa host galaxies as derived using SNIa data from \citet{Burns:2018ggj} (B18) and \citet{Riess:2016jrr} (R16). The Cepheid data set is the same in both cases. 
	\label{fig:magcompdiff}}
\end{figure}

The fact that the statistical error on $H_0$ is dominated by SNIa data, and only a sub-set of the SNIa are in tension with the Planck $H_0$ may raise concerns about the reliability of the individual SNIa magnitude estimates. However,  
using the SNIa data set from \citet{Burns:2018ggj}, gives very similar result, indicating that systematic effects related to the SNIa light curve fitting, stretch and color correction are subdominant, see figure~\ref{fig:magcompdiff} (assuming a fixed $R_{\rm E}=0.386$). The possibility that Cepheid calibrated SNIae to a larger extent originate in star-forming environments, thus being dimmer compared to the average Hubble flow SNIae, may bias the Hubble constant measurements has been discussed in \cite{Rigault_2015}. 

\section{Results Including Systematic Errors}\label{sec:results}
Finally, we update the LMC metallicity to $[{\rm M}/{\rm H}]=-0.4079$, rescale SNIa magnitude uncertainties, include metallicity uncertainties, as well as intrinsic color uncertainties and temperature variations by generating random Monte Carlo samples. Here, we use $k=-0.5\pm 0.25$, $Z_\odot=8.824\pm 0.15$. For approximate individual contributions to the error budget computed for the case of a fixed $R_{\rm E}=0.386$, see Table~\ref{tab:errorbudget}. The Hubble constant obtained in this case is $H_0 = 73.1\pm 1.8$ ($3.0\,\sigma$ Planck tension).

\begin{table}[!ht]
  \centering
  \caption{Approximate individual contributions to the systematic error budget for a fixed $R_{\rm E}=0.386$, with Int color being Cepheid intrinsic color uncertainties as given in \citet{Tammann_2011}, $\tau$ the intrinsic Cepheid temperature for which $\sigma_\tau=0.02$ \citep{Pejcha_2012}, $k$ and $Z_\odot$ relating metallicities as inferred from iron and oxygen abundances as described in Section~\ref{sec:metallicities}, and SNIa error being the effect of re-scaling SNIa magnitude uncertainties (see Section~\ref{sec:snmag}).} 
  \label{tab:errorbudget}
  \begin{tabular}{|c|c|}
    \hline
    Param  &$\sigma(H_0)/H_0$\\
    \hline
    Int color & $0.67\,\%$\\
    Int temperature $\tau$ & $0.41\,\%$\\
    Metallicity $k$ &  $0.34\,\%$\\
    Metallicity $Z_\odot$ & $1.2\,\%$\\
    SNIa error & $0.90\,\%$\\
    Total systematic & $1.37\,\%$\\
    \hline
  \end{tabular}
\end{table}

Allowing for $R_{\rm E}$ to vary between galaxies in the range $[0,1]$, we get $H_0 = 73.9\pm 2.2$ ($2.9\,\sigma$) with a GAIA residual parallax calibration offset of $zp=-19.3\pm 7.3\, \mu{\rm as}$. 
Assuming flat priors $R_{\rm E}=[0.15,0.8]$ we obtain $H_0 = 73.7\pm 2.1$ ($3.0\,\sigma$)
and for Gaussian priors $R_{\rm E}=0.48\pm 0.1$  $H_0 = 73.5\pm 2.1$ ($2.9\,\sigma$).

Calibrating MW Cepheid distances using companions and host cluster parallaxes following \citet{Breuval_2020}, we obtain $H_0 = 71.9\pm 2.2$ ($2.0\,\sigma$) assuming $R_{\rm E}=[0,1]$, with $zp=-11.0\pm 11.3\, \mu{\rm as}$, and $H_0 = 72.0\pm 2.2$ ($2.1\,\sigma$) and $H_0 = 72.4\pm 2.1$ ($2.4\,\sigma$) for 
$R_{\rm E}=[0.15,0.8]$ and $R_{\rm E}=0.48\pm 0.1$, respectively.

Applying a color excess cut, only using Cepheids for which $\hat{E}({\rm V} - {\rm I})<0.15$ mag, we obtain $H_0 = 68.3\pm 2.8$, for individually fitted $R_{\rm E}=[0,1]$, and $H_0 = 68.6\pm 2.5$ assuming a fixed global value of $R_{\rm E}=0.386$ following \citet{Riess:2016jrr}. Using the re-calibrated MW Cepheid data set from \citet{Breuval_2020}, imposing $\hat{E}({\rm V} - {\rm I})<0.15$ mag, the inferred Hubble constant is $H_0 = 68.1\pm 2.6$.

We thus note that mitigating the color calibration impact by removing Cepheids for which dust extinction is expected to dominate the observed color excess, the $H_0$ inferred from supernovae agree with the Planck value regardless of Cepheid color calibration parameterization. Cutting out the blue Cepheid tail has to the opposite, although less pronounced, effect of increasing the inferred $H_0$. Requiring $\hat{E}({\rm V} - {\rm I})>0$ mag, we obtain $H_0 = 74.8\pm 2.3$ for $R_{\rm E}=[0,1]$.

\section{Summary}\label{sec:summary}
\begin{figure}[!t]
    \centering
	\includegraphics[width=1\linewidth]{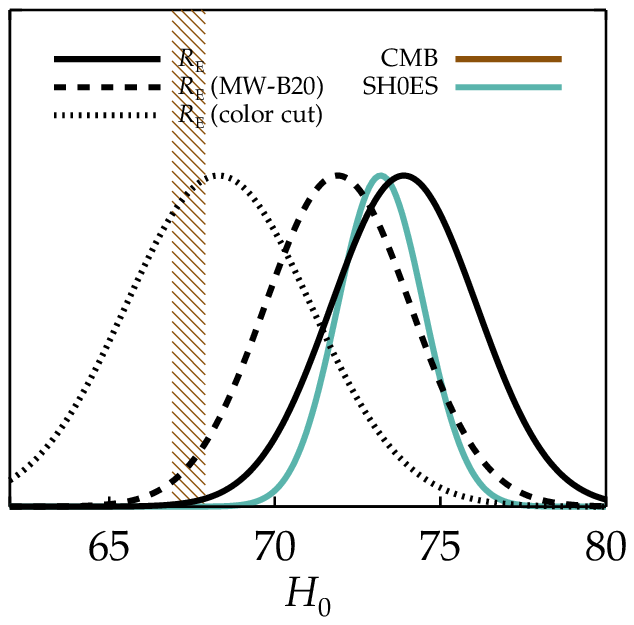}
	\caption{
	Comparing results for $H_0$. The solid black line is for individual galactic values of $R_{\rm E}=[0,1]$. For the dotted black line, we have imposed an upper limit on the allowed estimated color excess, only including Cepheids for which $\hat{E}\,({\rm V} - {\rm I})<0.15$ mag. For the dashed black line, we use the MW Cepheid sample from \citet{Breuval_2020}.
	The solid petrol line is fitted using the Wesenheit calibration with $R_{\rm W}= 0.386$ as in \citet{Riess_2021} and the dashed brown region indicates the $1\,\sigma$ region from Planck \citep{Planck2020}.
	\label{fig:Finalresult}}
\end{figure}

In \citet{mortsell2021hubble}, we investigated the sensitivity of the Hubble constant as inferred from SNIa distances to the choice of Cepheid color calibration method. Here, we complement this analysis by investigating the impact of Cepheid temperature variations, metallicity corrections, supernova magnitude uncertainties and Milky Way parallax systematics. The results are summarized in Figure~\ref{fig:Finalresult} and Table~\ref{tab:Finalresults}.

\begin{table}[!ht]
  \centering
  \caption{Summary of inferred $H_0$ for different calibration assumptions, including systematic effects. $R_{\rm E}=0.386$ assumes a fixed global value for Cepheid color color corrections. Ind gal $R_{\rm E}$ fits individual galactic values with flat priors $R_{\rm E}=[0,1]$ and $R_{\rm E}=[0.15,1]$ as well as Gaussian priors $R_{\rm E}=0.48\pm 0.1$. MW-B20 uses re-calibrated MW Cepheid distances following \citet{Breuval_2020}. The Red cut refers to only using Cepheids for which the intrinsic color $\hat{E}({\rm V} - {\rm I})<0.15$ mag. The Tension column refers to the tension with Hubble constant inferred from CMB observations with the Planck satellite \citep{Planck2020}.} 
  \label{tab:Finalresults}
  \begin{tabular}{|c|c|c|}
    \hline
    Calibration &$H_0$& Tension\\
    \hline
    $R_{\rm E}=0.386$ & $73.1\pm 1.8$ & $3.1\,\sigma$\\
    Ind gal $R_{\rm E}=[0,1]$ & $73.9\pm 2.2$ & $2.9\,\sigma$\\
    Ind gal $R_{\rm E}=[0.15,0.8]$ & $73.7\pm 2.1$ & $3.0\,\sigma$\\
    Ind gal $R_{\rm E}=0.48\pm 0.1$ & $73.5\pm 2.1$ & $2,9\,\sigma$\\
    MW-B20 $R_{\rm E}=[0,1]$ & $71.9\pm 2.2$ & $2.0\,\sigma$\\
    MW-B20 $R_{\rm E}=[0.15,0.8]$ & $72.0\pm 2.2$ & $2.0\,\sigma$\\
    MW-B20 $R_{\rm E}=0.48\pm 0.1$ & $72.4\pm 2.1$ & $2.4\,\sigma$\\
    Red cut MW-B20 $R_{\rm E}=[0,1]$ & $68.3\pm 2.8$ & $0.2\,\sigma$\\
    \hline
  \end{tabular}
\end{table}

Only using MW Cepheids for which distances can be estimated from companions and host cluster parallaxes \citep{Breuval_2020} and flat priors $R_{\rm E}=[0,1]$, we obtain $H_0 = 71.9\pm 2.2$, decreasing the difference with the Planck value to $2.0\,\sigma$, and in good agreement with $H_0=69.6\pm 1.6$ obtained calibrating the absolute SNIa magnitude using the tip of the red giant branch observations \citep{Freedman_2019}.
Using more restricted prior values for $R_{\rm E}$ yield similar results, see Table~\ref{tab:Finalresults}.

A possible caveat with ours, and all other $H_0$ measurements using Cepheid calibrated distances, comes from the fact that the observed Cepheid color excess distribution have large variations across galaxies, and that the inferred Hubble constant is sensitive to color excess cuts in the data. Also applying a color excess cut to remove Cepheids for which
dust extinction is expected to dominate the observed color excess,
$\hat{E}({\rm V} - {\rm I})>0.15$ mag, we obtain $H_0 = 68.1\pm 2.6$.

We thus conclude that the current Hubble tension may be affected by systematic effects in the calibration of the local distance ladder, including MW Cepheid distances, as well as Cepheid selection biases.

\begin{acknowledgements}
We thank Vallery Stanishev for interesting discussions regarding stellar metallicities and the anonymous referee for useful comments on the manuscript.
EM acknowledges support from the Swedish Research Council under Dnr VR 2020-03384.
AG acknowledges support from the Swedish Research Council under Dnr VR 2020-03444, and the Swedish National Space Board, grant 110-18.
\end{acknowledgements}

\bibliography{bibliography}{}
\bibliographystyle{aasjournal}



\end{document}